\documentclass[11pt,a4paper]{article}
\usepackage{jcappub}
\usepackage{graphicx}	
\usepackage{amsmath, cases,mathtools,tensor}
\usepackage{caption}
\usepackage{subcaption}
\newcommand{\F}{\mathbb{F}}

\newcommand{\R}{\mathbb{R}}

\title{Decoupled Gravitational Wave Equations in Spherical Symmetry from Curvature Wave Equations}

\author{Gowtham Rishi Mukkamala and David Pereñiguez}

\affiliation{Niels Bohr International Academy, Niels Bohr Institute, Copenhagen University,\\
Blegdamsvej 17, DK-2100 Copenhagen, Denmark}

\emailAdd{glq382@alumni.ku.dk}
\emailAdd{david.pereniguez@nbi.ku.dk}

\abstract{Black hole perturbation theory on spherically symmetric backgrounds has been instrumental in establishing various aspects about the gravitational dynamics close to black holes, and continues to be an interesting avenue to confront current challenges in gravitational physics. In this paper, we present an approach to perturbation theory in spherical symmetry that addresses simultaneously some conceivably inconvenient aspects of the traditional methods. In particular, focusing on Schwarzschild's background we are able to derive a decoupled wave equation, for a single complex variable, by simply computing one component of the curvature wave equation satisfied by a complex self-dual version of the Riemann tensor. The real and imaginary parts of the variable consist only of even and odd pieces of the metric fluctuation, respectively, and both satisfy the Regge--Wheeler equation. Besides providing a systematic derivation of decoupled equations, an immediate corollary of our results is the isospectrality between even and odd sectors. We conclude by discussing potential extensions of our formalism to include matter and higher orders in perturbation theory.}

\begin{document}
\maketitle
\clearpage
\section{Introduction} \label{Intro}

The assumption of spherical symmetry has proved being extremely powerful throughout the history of General Relativity (GR), as it allows one to extract very valuable lessons about gravitating systems without facing the often major difficulties intrinsic to more realistic situations. The most basic and celebrated instance of this is the early finding of Schwarzschild's solution \cite{SchwarzschildKarl}, which lead to the theoretical discovery and significant understanding of black holes well before the most general (stationary and vacuum) solution was found by Kerr \cite{PhysRevLett.11.237}. Another important example is the study of gravitational dynamics in the vicinity of black holes. Since its early days \cite{PhysRev.108.1063,PhysRevLett.24.737}, black hole perturbation theory about spherically symmetric solutions has been instrumental in elucidating the dynamical behaviour of gravitation in the strong field regime. Amongst its numerous applications, it allowed theorists to establish that black holes relax towards equilibrium in a \textit{ringdown} process, consisting of a superposition of so-called quasinormal modes (QNMs) with some characteristic frequencies and damping times that depend only on the black hole parameters (see e.g.~\cite{Berti:2009kk} and references therein).\footnote{In fact, perturbation theory also predicts that following the QNM ringing there is an inverse power-law behaviour in time, or tails \cite{PhysRevD.5.2419}, where the precise value of the exponent is also fixed by the theory.} Following the first groundbreaking detection of a black hole merger \cite{LIGOScientific:2016aoc}, and due to its enormous discovery potential, the understanding and precise measurement of the black hole's ringdown has become a top priority in gravitational physics.

The original approaches by Regge, Wheeler and Zerilli (RWZ) \cite{PhysRev.108.1063,PhysRevLett.24.737} were superseded by Teukolsky's \cite{Teukolsky:1973ha}, in the sense that the latter allows one to also account for rotating black holes. Nevertheless, it is reasonable to expect that the assumption of spherical symmetry will be as effective in tackling the emerging challenges in modern gravitational physics as it has been in addressing those of the past. From that lens, it is apparent that developing further our understanding of black hole perturbation theory in spherical symmetry is an important task. A motivation of relevance nowadays is understanding the non-linearities present in the ringdown of a black hole (such as quadratic modes \cite{Gleiser:1996yc,Nicasio:1998aj,Zlochower:2003yh,Mitman:2022qdl,Cheung:2022rbm,Ma:2022wpv,Redondo-Yuste:2023seq,Cheung:2023vki,Qiu:2023lwo,Khera:2023oyf}, absorption-induced mode excitations \cite{Sberna:2021eui,Redondo-Yuste:2023ipg,May:2024rrg,Zhu:2024dyl}, unconventional tails \cite{Okuzumi:2008ej,PhysRevD.108.084037,Carullo:2023tff,Cardoso:2024jme,DeAmicis:2024not,Islam:2024vro}, or even turbulence \cite{Bizon:2011gg,Dias:2011ss,Green:2013zba,Yang:2014tla}, see also \cite{Santos-Olivan:2015yok,Santos-Olivan:2016djn}), whose detectability with gravitational wave facilities has been recently assessed yielding exciting prospects \cite{Yi:2024elj}. Such non-linearities can be studied using higher order-perturbation theory which is, in principle, technically more approachable in spherical symmetry \cite{Gleiser:1995gx,Gleiser:1998rw,Garat:1999vr,Nakano:2007cj,Brizuela:2006ne,Brizuela:2007zza,Brizuela:2009qd,Ioka:2007ak,Bucciotti:2023ets,Perrone:2023jzq,Bucciotti:2024zyp,Bucciotti:2024jrv,BenAchour:2024skv,Spiers:2023mor,Riva:2023rcm} than in Kerr \cite{Campanelli:1998jv,Green:2019nam,Loutrel:2020wbw,Ripley:2020xby,Spiers:2023cip,Ma:2024qcv,Bourg:2024jme}. Given that the structure of \textit{linear} perturbation theory is nested into all next perturbative orders (see e.g.~\cite{Lagos:2022otp} for a detailed discussion), one hopes that improvements at first order carry over to the higher orders, too.

The last decades have seen significant progress added to the pioneering works by RWZ. These include the gauge-invariant approach by Moncrief \cite{Moncrief:1974am,PhysRevD.12.1526}, the covariant formalisms introduced by Gerlach and Sengupta \cite{Gerlach:1979rw,Gerlach:1980tx}, as well as further developments that involve the inclusion of certain types of matter \cite{Gundlach:1999bt,Martin-Garcia:2000cgm,Cardoso:2001bb,Martel:2005ir,Chaverra:2012bh,Pereniguez:2023wxf,Speeney:2024mas}, higher dimensions \cite{Kodama:2000fa,Kodama:2003jz,Ishibashi:2011ws} or the study of the underlying symmetries of the gravitational equations \cite{Lenzi:2021wpc,Lenzi:2021njy,Solomon:2023ltn,Jaramillo:2024qjz,Lenzi:2024tgk}, amongst other works. In spite of this there are, in our view, a few aspects that could still be clarified further. On the one hand, the derivations of the master wave equations governing the even and odd sectors of the fluctuation are quite non-systematic, and rely heavily on, a priory, arbitrary combinations of variables and equations (this was discussed in detail in \cite{Lenzi:2021wpc,Lenzi:2024tgk}). It would thus be desirable to have decoupled wave equations whose geometric origin is clear. On the other hand, while the potentials of the even and odd master equations are quite different, as is well known they turn out to be isospectral, that is, they share the same characteristic quasinormal frequencies \cite{Chandrasekhar:1985kt}. This suggests that an equal footing treatment should be possible where statements like isospectrality follow immediately (and all of it without resorting to perturbation theory methods based on the Petrov type of the spacetime instead of on spherical symmetry).

In this work, we present an approach that addresses both issues simultaneously. The starting point is the non-perturbative wave equation satisfied in vacuum by a complex self-dual version of the Riemann curvature tensor, $\mathbb{R}_{\mu\nu\rho\sigma}$, that we refer to as a \textit{curvature wave equation}. Upon linearisation on Schwarzschild's background, and working mostly in terms of the components of $\delta \mathbb{R}_{\mu\nu\rho\sigma}$ instead of the metric perturbation $h_{\mu\nu}$, we are able to derive a decoupled wave equation, for a single complex variable, by just computing one component of the linearised curvature wave equation of $\mathbb{R}_{\mu\nu\rho\sigma}$. The complex variable is simply a linear combination of one component of $\delta \mathbb{R}_{\mu\nu\rho\sigma}$, denoted $\mathcal{I}$, and another of $h_{\mu\nu}$, and its real and imaginary parts (after writing $\mathcal{I}$ in terms of $h_{\mu\nu}$) involve only even and odd pieces of $h_{\mu\nu}$, respectively. Given that the differential operator and the potential of the equation are real, isospectrality follows as a direct consequence.\footnote{The QNM boundary conditions, i.e. regularity at the horizon and outgoing at infinity, are also the same for both real and imaginary parts of the variable.} In addition, the potential is precisely that found by Regge and Wheeler, which is significantly simpler than Zerilli's, and in our case it is the one governing the dynamics of \textit{both} sectors (while this may come as a surprise, we note that a similar result was found previously by \cite{Sasaki1981TheRE,Chaverra:2012bh}). While curvature wave equations have been used in perturbation theory within the Newman--Penrose formalism (the pioneering work was by Ryan \cite{Ryan:1974nt}, which was later extended and employed by others, see e.g.~\cite{Campanelli:1998jv,Bini:2002jx,Aksteiner:2010rh}), we stress that our approach here is new and entirely based on the natural structures of spherical symmetry.\footnote{From a different approach, curvature wave equations were also used to derive master equations in spherically symmetric black holes in higher-dimensions in \cite{Godazgar:2011sn,Cai:2013cja}.} We also note that our results are markedly four-dimensional, although some may be extendable to higher-dimensions in certain cases, as we shall argue. Finally, since the formalism can be extended straightforwardly to electromagnetism, we use that as a warm up example preceding the gravitational case. 

Our paper is organised as follows. In Section \ref{Sec1} we motivate the introduction of the complex self-dual curvature tensors for electromagnetism and gravity, and provide their non-perturbative curvature wave equations in vacuum. In Section \ref{Sec2} we discuss the linearisation of such tensors, their curvature wave equations, and derive the respective decoupled master equations. We conclude in Section \ref{Sec3} by summarising and discussing our results.

\section{Self-dual Curvature Tensors and Their Wave Equations} \label{Sec1}

An unavoidable consequence of Special Relativity is that the propagation of any physical degrees of freedom must be in the form of waves. In covariant gauge theories such degrees of freedom are encoded in the curvature tensors, but the equations of motion as derived form a least action principle are often not in the form of a wave equation for the curvature (e.g. $\nabla^{\mu}F_{\mu\nu}=0$ or $R_{\mu\nu}=0$). Therefore, it must be possible to write some notion of \textit{curvature wave equation}. The purpose of this section is to derive such equations, in vacuum, for the gravitational and the electromagnetic fields. Although this has been addressed in several occasions in the literature (as well as generalisations, see \cite{Bini:2003km,Edgar:2005zr}), our discussion is closest in spirit to \cite{Bini:2002jx}. Note that the equations in this section are fully general, and rely on neither symmetry assumptions nor perturbative expansions.

\subsection{Electromagnetism} \label{Sec1.1}

The equations of motion for an electromagnetic field evolving on a curved background read, in the absence of sources,\footnote{The exterior derivative and Hodge dual of a differential $p$-form $X_{\mu_{1}...\mu_{p}}$ are $dX_{\mu_{1}...\mu_{d+1}}=(p+1)\nabla_{[\mu_{1}}X_{\mu_{2}...\mu_{p+1}]}$ and $\star X_{\mu_{1}...\mu_{d-p}}=(1/p!)\epsilon_{\mu_{1}...\mu_{d-p}\nu_{1}...\nu_{p}}X^{\nu_{1}...\nu_{p}}$, where $\nabla_{\mu}$ is the metric covariant derivative and $\epsilon_{\mu_{1}...\mu_{d}}$ is the metric volume form.}
\begin{equation}\label{F1}
    d F_{\mu\nu\rho}=0\, , \quad \quad d \star F_{\mu\nu\rho}=0\,,
\end{equation}
where $F_{\mu\nu}=2\partial_{[\mu}A_{\nu]}$ is Maxwell's field strength. It is well known that Maxwell's equations enjoy invariance under electric-magnetic $SO(2)$ duality transformations. In terms of the complex field strength,
\begin{equation}\label{F2}
    \F_{\mu\nu} \coloneqq F_{\mu\nu}-i\star F_{\mu\nu}\, , 
\end{equation}
such transformations read simply $\F\mapsto e^{i\alpha}\F$, where $\alpha$ is a constant. If one also considered the coupling to gravity, employing \eqref{F2} in perturbation theory makes it easy to keep duality-invariance manifest at all times and, following the observation in \cite{Pereniguez:2023wxf}, we will see that it is working in terms of \eqref{F2} and not $F_{\mu\nu}$ what allows a treatment in equal footing between even and odd sectors (see \cite{Jaramillo:2023day} for other observations in favour of employing self-dual variables). In Section \ref{Sec2} we will show that an analogous statement holds for gravity, too. The equations of motion \eqref{F1} and the definition \eqref{F2} imply 
\begin{equation}\label{F3}
    d\F_{\mu\nu\rho}=0\, ,  \quad \quad \F_{\mu\nu}+i\star \F_{\mu\nu}=0\,, 
\end{equation}
and we say that $\F$ is \textit{self-dual} in the sense that it satisfies the second of the equations above. Next, we need a second order operator acting in the space of self-dual curvature tensors. A natural choice in Maxwell's theory is the usual harmonic operator
\begin{equation}
    \triangle\coloneqq d\star d\star+\star d\star d \, .
\end{equation}
Then, equations \eqref{F3} imply
\begin{equation}\label{triangleF}
    \triangle\F_{\mu\nu}=0\, ,
\end{equation}
which we can use as the curvature wave equation for Maxwell's field. Alternatively, in terms of the spacetime covariant derivative $\nabla_{\mu}$, and after a straightforward use of Ricci's identity, \eqref{triangleF} can be recast in the form
\begin{equation}
 \square \F_{\mu\nu}+2 \tensor{R}{_{[\mu}^\gamma}\tensor{\F}{_{\nu]}_\gamma}+\tensor{R}{_\mu_\nu^\alpha^\gamma}\F_{\alpha\gamma}=0\, ,
\end{equation}
where $\square=\nabla^{\mu}\nabla_{\mu}$ is d'Alembert's operator. As we will show in Section \ref{Sec2}, the master wave equation governing fluctuations of the electromagnetic field is obtained by computing just one component of the linearised curvature equation $\delta(\triangle \F)_{\mu\nu}=0$.

\subsection{Gravity}  \label{Sec1.2}

In order to emulate the steps followed in the electromagnetic case, we should interpret the Riemann tensor $\tensor{R}{_\mu_\nu_\alpha_\beta}$ as a curvature 2-form. This is a particular case of tensor-valued differential $p$-forms, which are tensor fields with index structure $X_{\mu_{1}...\mu_{p}\mathcal{A}}$, where $\mathcal{A}$ denotes any index array, such as $\mathcal{A}=\alpha$, $\mathcal{A}=\alpha\beta$, etc, and is fully skew-symmetric in $\mu_{1}...\mu_{p}$, or in other words, it has the property $X_{\mu_{1}...\mu_{p}\mathcal{A}}=X_{[\mu_{1}...\mu_{p}]\mathcal{A}}$.\footnote{We should stick to the convention of writing first the skew-symmetric indices of a tensor-valued differential $p$-form.} Notions such as the Hodge dual or the exterior derivative of ordinary differential forms are generalised to tensor-valued $p$-forms as\footnote{However, we recall that the exterior covariant derivative $D$ lacks the property $d^{2}=0$ of the usual exterior derivative, that is, $D^{2}\ne0$.}
\begin{align}
    \star X_{\mu_{1}...\mu_{d-p}\mathcal{A}}\coloneqq& \frac{1}{p!} \tensor{\epsilon}{_{\mu_{1}}_{...}_{\mu_{d-p}}^{\nu_{1}}^{...}^{\nu_{p}}} \tensor{X}{_{\nu_{1}}_{...}_{\nu_{p}}_{\mathcal{A}}}\, ,\\ \notag \\
    D_{\mu}\tensor{X}{_{\mu_{1}}_{...}_{\mu_{p}}_{\mathcal{A}}}\coloneqq&(p+1)\nabla_{[\mu}\tensor{X}{_{\mu_{1}}_{...}_{\mu_{p}]}_{\mathcal{A}}}\, .
\end{align}
With this operators at hands, and from the perspective that $\tensor{R}{_\mu_\nu_\alpha_\beta}$ is a tensor-valued differential 2-form, we can reproduce the steps followed in the electromagnetic case to arrive at a curvature wave equation. 

The Einstein equation in vacuum is 
\begin{equation}\label{eqRic}
    R_{\mu\nu}=0\, ,
\end{equation}
which, quite crucially, implies that the Hodge duals of the Riemann tensor as taken from the left and from the right coincide, that is, $\tensor{\epsilon}{_\mu_\nu^\alpha^\beta}R_{\alpha\beta\sigma\rho}=\tensor{\epsilon}{_\sigma_\rho^\alpha^\beta}R_{\alpha\beta\mu\nu}$ (see e.g. \cite{Penrose:1985bww}). Thus, we can unambiguously introduce the self-dual complex curvature tensor
\begin{equation}\label{CompR}
    \R_{\mu\nu\rho\sigma}\coloneqq R_{\mu\nu\rho\sigma}-i \star R_{\mu\nu\rho\sigma}\, ,
\end{equation}
which in vacuum inherits the algebraic symmetries of the Riemann tensor. In gravity there is no equivalent of the electric-magnetic duality of electromagnetism, with the exception of linearised gravity on Minkowski's space \cite{Henneaux:2004jw}. On the one hand, working with \eqref{CompR} allows one to keep such symmetry explicit whenever it is present but, most importantly for us, employing \eqref{CompR} will yield a treatment on equal footing between the even and odd sectors of the fluctuation. The definition \eqref{CompR} and equation \eqref{eqRic} imply that\footnote{We notice that while $D_{\mu} R_{\nu\rho\sigma\delta}=0$ is an identity, one has to use $R_{\mu\nu}=0$ to show that $D_{\mu}\star R_{\nu\rho\sigma\delta}=D_{\mu} R\star_{\nu\rho\sigma\delta}=(1/2)\epsilon^{\alpha\beta}_{\ \ \ \sigma\delta}(D_{\mu} R_{\nu\rho\alpha\beta})=0$. In parallel, from the symmetry $R_{\alpha[\mu\nu\rho]}=0$ it is immediate that $\R^{\alpha}_{\ \mu\alpha\nu}=R_{\mu\nu}=0$.}
\begin{equation}\label{DR}
    D_{\alpha}\R_{\mu\nu\rho\sigma}=0\, , \quad \quad \R_{\mu\nu\rho\sigma}+i\star \R_{\mu\nu\rho\sigma}=0 \, , \quad \R^{\alpha}_{\ \mu\alpha\nu}=0\, ,
\end{equation}
and, similarly to the electromagnetic case, as a second order operator acting in the space of self-dual curvature forms for gravity we can take 
\begin{equation}
    \triangle\coloneqq D\star D\star+\star D\star D\, .
\end{equation}
Then, equations \eqref{DR} imply
\begin{equation}\label{triangle}
    \triangle \R_{\mu\nu\rho\sigma}=0\, ,
\end{equation}
which we can use as the curvature wave equation for the gravitational field. By employing Ricci's identity and using \eqref{eqRic} this equation can be written as
\begin{equation}\label{GravCWE}
    \square \R_{\mu\nu\rho\sigma} +\tensor{R}{_\mu_\nu^\gamma^\lambda}\R_{\gamma\lambda\rho\sigma}+2\left(\tensor{R}{_\rho^\lambda_{[\mu}^\gamma}\R_{\nu]\gamma\sigma\lambda}-\tensor{R}{_\sigma^\lambda_{[\mu}^\gamma}\R_{\nu]\gamma\rho\lambda}\right)=0\, .
\end{equation}
Closely resembling the electromagnetic case, the master wave equation for the gravitational field will follow by computing just one component of the linearised curvature equation $\delta\left( \triangle \R\right)_{\mu\nu\rho\sigma}=0$.

\section{Decoupled Wave Equations} \label{Sec2}

In this section we use the curvature wave equations presented above to derive decoupled wave equations governing generic linear fluctuations about spherically symmetric solutions. We will be using the covariant approach by \cite{Gerlach:1979rw,Gerlach:1980tx}, developed further in e.g. \cite{Martel:2005ir} (in particular, we will adopt the conventions in \cite{Pereniguez:2023wxf}). Since this is a fairly well-known formalism, here we will just review the basic ideas and refer the reader to the previous references for the details. The starting point is a background space with manifold structure $N^{2}\times S^{2}$ and metric, 
\begin{equation}
    ds^{2}=g_{ab}(y)dy^{a}dy^{b}+r^2(y)\Omega_{AB}(z)dz^{A}dz^{B}\, ,
\end{equation}
where $g_{ab}(y)$, and $r^{2}(y)$ are a Lorentzian metric and a function in $N^{2}$, which is a 2-dimensional manifold parameterised by the coordinates $y^{a}$ (with $a=1,2$), and $\Omega_{AB}(z)$ is the round metric on the 2-sphere $S^{2}$, parametrised by generic coordinates $z^{A}$ (with $A=3,4$). Next, the idea is to expand the fluctuations of a quantity $\delta \mathcal{Q}$ in even and odd spherical harmonics, schematically,\footnote{Here the subscripts ``e'' and ``o'' stand for even and odd, respectively. We recall that our conventions for the spherical harmonics are those in \cite{Pereniguez:2023wxf}.}
\begin{equation}
   \delta \mathcal{Q}=\sum_{l,m}q^{(l,m)}_{\text{e,o}}(y)Y_{\text{e,o}}^{(l,m)}(z)\,,
\end{equation}
and reduce the equations of motion to PDEs for $q^{(l,m)}_{\text{e,o}}(y)$ in $N^{2}$. Although this reduction to $N^{2}$ is very useful, the system of PDEs is still rather involved, and its reduction into decoupled wave equations is quite unilluminating. Below we will show that such decoupled equations simply follow from a component of the curvature wave equations. For concreteness, in this work we will focus on Schwarzschild's background and consider only the radiative modes of electromagnetic ($l\geq1$) and gravitational ($l\geq2$) fluctuations, which are decoupled at first order and can thus be studied separately. Then, one can always work in the Regge-Wheeler gauge or, alternatively, in terms of the associated gauge-invariant quantities (see \cite{Gerlach:1979rw,Gerlach:1980tx,Martel:2005ir,Pereniguez:2023wxf}). An extension of our approach to include the coupling of gravity to arbitrary matter (where $l=0,1$ modes may also be radiative) will be given in subsequent works.

Before addressing the computation, it is worth noticing that our spherical harmonic tensors (as defined in \cite{Pereniguez:2023wxf}) are complex-valued and have the property $Y_{\text{e,o}}^{l,-m}=(-1)^{m}\bar{Y}_{\text{e,o}}^{l,m}$, where a bar denotes complex conjugation. Given that we will expand quantities that are complex-valued (such as the self-dual curvature tensors), it is useful to introduce the ``real'' and ``imaginary'' parts of the harmonic components of a fluctuation $\delta \mathcal{Q}$, as 
\begin{equation}\label{pm}
    \left(q^{(l,m)}_{\text{e,o}}\right)^{\pm}\coloneqq q^{(l,m)}_{\text{e,o}}\pm(-1)^{m}\bar{q}^{(l,-m)}_{\text{e,o}}\, .
\end{equation}
Then, if $\delta \mathcal{Q}$ is real (like $h_{\mu\nu}$) its harmonic components satisfy $\left(q^{l,m}_{\text{e,o}}\right)^{+}=2q^{l,m}_{\text{e,o}}$ and $\left(q^{l,m}_{\text{e,o}}\right)^{-}=0$. However, this is not necessarily true if $\delta \mathcal{Q}$ is complex-valued (like $\delta \R_{\mu\nu\rho\sigma}$). Finally, to avoid cluttering our expressions we will omit writing explicitly the harmonic labels $(l,m)$, but summation over harmonics should always by assumed (unless otherwise stated). 

\subsection{Electromagnetism } \label{Sec2.1D}

The starting point is defining the harmonic components of the perturbation. Instead of expanding Maxwell's vector potential $\delta A_{\mu}$, we shall follow \cite{Pereniguez:2023wxf} and work in terms of $\delta \F_{\mu\nu}$. Its expansion in harmonics reads
\begin{equation}\label{EM Harmonic Expansion}
    \delta \mathbb{F} = \frac{1}{2} A(y) Y \epsilon_{ab}  dy^{a}  \wedge dy^{b} + \frac{1}{2} \Phi (y) Y \epsilon_{AB} dz^{A} \wedge dz^{B} 
    + (B_{a}(y) Z_{A} + C_{a}(y) X_{A} ) dy^{a}  \wedge dz^{A},
\end{equation}
where $Y$ denotes the shperical harmonics, $Z_{A}$ and $X_{A}$ are the even and odd harmonics vectors and $\epsilon_{ab}$ and $\epsilon_{AB}$ denote the volume forms on $N^{2}$ and the sphere, respectively. $A(y), \Phi (y), B_{a}(y) $ and $C_{a}(y)$ are tensors on $N^{2}$ subject to a system of PDEs which follow after plugging \eqref{EM Harmonic Expansion} into the linearised Maxwell equations \eqref{F1}. By judicious manipulations of those equations, exploiting the decoupling between even and odd sectors, one can get to a decoupled wave equation involving a single quantity per sector (see e.g.\cite{PhysRevD.12.1526,Pereniguez:2023wxf}). Here, we instead show that a decoupled wave equation is obtained directly by proceeding as follows:
\begin{enumerate}
    \item First, we compute the single purely spherical component of the linearised curvature wave equation \eqref{triangleF}, that is,
    \begin{equation}\label{triF}
        \epsilon^{AB}\left(\delta\triangle \F\right)_{AB}=0\, .
    \end{equation}
    \item Next, we use the linearised one-derivative equation \eqref{F3},
    \begin{equation}
    \delta\left( d\F \right)_{\mu\nu\rho} =0\, ,
    \end{equation}
    to perform a straightforward elimination of variables to write \eqref{triF} as an equation for $\Phi(y)$ only. 
    \item In the steps above, we use the background equations satisfied by the Schwarzschild solution,
    \begin{align}\label{bg}
     r_{a}r^{a} = 1 - \frac{2 M}{r}\, ,\quad \nabla_{a}r_{b} = \frac{M g_{ab}}{r^2}\, ,
    \end{align}
    to substitute $r^{a}r_{a}$ and the gradient $\nabla_{a}r_{b}$, where $\nabla_{a}$ denotes the covariant derivative of the Lorentzian metric $g_{ab}(y)$, $r_{a}=\nabla_{a}r$ and $M$ is the ADM mass.
\end{enumerate}
Indeed, by symmetry the purely spherical part of $\delta\triangle\F $ satisfies $\left(\delta\triangle \F\right)_{AB}\sim \epsilon_{AB}$, so it has a single independent component. Computing it one finds the equation
\begin{equation}
     \epsilon^{AB}\left(\delta\triangle \F\right)_{AB}= \left[\nabla_{a}\nabla^{a}\Phi-\frac{2 (\nabla_{a} r)(\nabla^{a} \Phi ) }{r} - l(l+1)\left(\frac{\Phi }{r^2} - \frac{2 C^{a}\nabla_{a}r}{r } \right)  \right]Y=0\,.
\end{equation}
Then, $C_{a}$ can be easily traded by $\Phi$ noticing that
\begin{equation}
    \epsilon^{AB}(\delta d \F)_{aAB}=\left[\nabla_{a}\Phi -l(l+1) C_{a} \right]Y=0\, ,
\end{equation}
and one gets a decoupled equation for $\Phi$,
\begin{align}\label{EMmasterwaveeq}
    \Box \Phi - \frac{l(l+1)}{r^2} \Phi  = 0\, ,
\end{align}
where $\Box=\nabla^{a}\nabla_{a}$ is the d'Alembertian of $g_{ab}(y)$ on $N^{2}$. On the one hand, this is a straightforward and systematic derivation of a decoupled equation for $\delta \F_{\mu\nu}$. On the other hand, being a component of a complex-valued quantity, the complex variable $\Phi$ describes the degrees of freedom of both traditional even and odd sectors. To see this, we follow \cite{Pereniguez:2023wxf} and expand the fluctuation of the real-valued Maxwell field strength as
\begin{align}\notag
\delta F=\frac{1}{2!}\mathcal{E}(y)Y\epsilon_{ab}dy^{a}\wedge dy^{b}+\frac{1}{2!}\mathcal{B}(y)Y\epsilon_{AB}dz^{A}\wedge dz^{B}+\Big(\mathcal{E}_{a}(y)Z_{A}+\mathcal{B}_{a}(y)X_{A}\Big)dy^{a}\wedge dz^{A}\, ,
\end{align}
where the tensors $\mathcal{E}(y),\mathcal{E}_{a}(y)$ and $\mathcal{B}_{a}(y),\mathcal{B}(y)$ correspond to the even and odd sectors and are related to the electric and magnetic field fluctuations, respectively. Writing the $+$ and $-$ components of $\Phi$ in terms of those (see \eqref{pm}), one finds 
\begin{equation}\label{pmEle}
    \Phi^{+}=2\mathcal{B}(y)\, ,\quad \Phi^{-}=2ir^{2}\mathcal{E}(y)\, .
\end{equation}
That is, $\Phi^{\pm}$ consist purely of either even or odd variables and satisfy the \textit{same} wave equation \eqref{EMmasterwaveeq} (given that both the operator $\Box$ and the potential are real and independent of the harmonic azimutal number $m$). In particular, the quasinormal mode boundary conditions are the same for both $\mathcal{B}(y)$ and $\mathcal{E}(y)$ (that is, regularity at the horizon and purely outgoing at future null infinity), so they also satisfy the same characteristic value problem for the quasinormal frequencies. Or in other words, an immediate consequence of \eqref{EMmasterwaveeq} and \eqref{pmEle} is the isospectrality between the even and odd sectors of an electromagnetic fluctuation. Remarkably, as we show below, this procedure can be reproduced closely in the gravitational case, leading to analogous results.

\subsection{Gravity }

In vacuum ($R_{\mu\nu}=0$) all algebraic symmetries of the Riemann tensor are inherited by $\R_{\mu\nu\rho\sigma}$. Therefore, the independent components of its linearisation $\delta \R_{\mu\nu\rho\sigma}$ can be written as
\begin{equation}\label{RiemannHarmonicProjD}
    \begin{split}
        \delta \mathbb{R}_{abcd} &= I Y  {\epsilon}_{ab} {\epsilon}_{cd},\\
        \delta \mathbb{R}_{Abcd} &= \left(\mathcal{U}_b Z_A + \mathcal{J}_b X_A \right){\epsilon}_{cd},\\
        \delta \mathbb{R}_{AaBb} &= \mathcal{A}_{ab}Y {\epsilon}_{AB} + \mathcal{S}_{ab} U_{AB} + \mathcal{H}_{ab} V_{AB} + \Upsilon_{ab} W_{AB},\\
        \delta \mathbb{R}_{abAB} &= \mathcal{N} Y {\epsilon}_{ab}  {\epsilon}_{AB},\\
        \delta \mathbb{R}_{ABDe} &=\left( \mathcal{F}_e Z_D + \mathcal{G}_e X_D\right) {\epsilon}_{AB},\\
        \delta \mathbb{R}_{ABCD} &= \mathcal{I} Y {\epsilon}_{AB} {\epsilon}_{CD},
    \end{split}
\end{equation}
where $I,\mathcal{N},\mathcal{I},\mathcal{U}_b, \mathcal{J}_b,\mathcal{A}_{ab}, \mathcal{S}_{ab}, \mathcal{H}_{ab}, \Upsilon_{ab}, \mathcal{F}_e,$ and $\mathcal{G}_e$  are tensors on $N^{2}$. These are, however, constrained further by the symmetries of $\delta \R_{\mu\nu\rho\sigma}$. First, $\delta \mathbb{R}_{\mu \nu \alpha \beta} = \delta \mathbb{R}_{\alpha \beta \mu \nu }$ implies that $\mathcal{A}_{ab}$ is skew-symmetric whereas tensors $\mathcal{S}_{ab}, \mathcal{H}_{ab} $ and $\Upsilon_{ab}$ are symmetric. In turn, from $\mathbb{R}_{[\mu \nu \alpha] \beta} = 0$ one finds that
\begin{equation}\label{AlgeBianchiD}
    \mathcal{A}_{ab} = \frac{\mathcal{N}}{2} \epsilon_{ab}\, .
\end{equation}
An important point of our derivation is that we do \textit{not} substitute the curvature components \eqref{RiemannHarmonicProjD} by their expressions in terms of the metric fluctuation (this is kept implicit at all times). The latter, in the Regge-Wheeler gauge (or equivalently the associated gauge-invariant variable) reads, in the conventions of \cite{Pereniguez:2023wxf},
\begin{equation}\label{met}
h=h_{ab}Ydy^{a}dy^{b}+2j_{a}X_{A}dy^{a}dz^{A}+k U_{AB}dz^{A}dz^{B}\, ,
\end{equation}
with $h_{ab}(y),j_{a}(y)$ and $k(y)$ tensors on $N^{2}$, and $U_{AB}=\Omega_{AB}Y$. With this we are in conditions to proceed analogously to the electromagnetic case:
\begin{enumerate}
    \item First, we compute the purely spherical component of the linearised curvature wave equation \eqref{triangle}, whose left hand side by symmetry must have the structure $(\delta\triangle\R)_{ABCD}\sim\epsilon_{AB}\epsilon_{CD}$,
    \begin{equation}\label{delR}
        \epsilon^{AB}\epsilon^{CD}(\delta\triangle\R)_{ABCD}=0\, .
    \end{equation}
    As mentioned above, in writing this equation we do not substitute the curvature components \eqref{RiemannHarmonicProjD} by their expression in terms of the metric perturbation \eqref{met}.\footnote{Of course, in spite of this in the linearisation \eqref{delR} there will still be explicit dependencies in \eqref{met}, but none coming from $\delta \R_{\mu\nu\rho\sigma}$.}
    \item Next, we use the linearised zero- and one-derivative equations for the curvature \eqref{DR}, 
    \begin{equation}\label{deltasR}
        \delta \left(D_{\alpha}\R_{\mu\nu\rho\sigma}\right)=0\,, \quad \delta\left(\R_{\mu\nu\rho\sigma}+i\star \R_{\mu\nu\rho\sigma}\right)=0\, , \quad \delta \left(\R^{\alpha}_{\ \mu\alpha\nu}\right)=0\, ,
    \end{equation}
    to make straightforward substitutions and write equation \eqref{delR} as an equation for $\mathcal{I}$ and $k$ only. 
    \item As in the electromagnetic case, in the two steps above we employ the background equations \eqref{bg} to replace the gradients of $r_{a}$ and the products $r^{a}r_{a}$.
\end{enumerate}
Computing \eqref{delR}, one finds an expression of the form
\begin{equation}\label{premast}
    \epsilon^{AB}\epsilon^{CD}(\delta\triangle\R)_{ABCD}=A_{1}\left(\mathcal{I}, \mathcal{N},\mathcal{G}_{a},\mathcal{S}_{ab}, \mathcal{A}_{ab}; k,j_{a} ,h_{ab}\right)Y+A_{2}(\Bar{\mathcal{S}}_{ab},\Bar{\mathcal{A}}_{ab}) \Bar{Y}=0\, .
\end{equation}
Although the exact form of this equation is still uninformative (we provide it explicitly in Appendix \ref{App}), that will be no problem since substantial cancellations take place automatically after straightforward substitutions using \eqref{deltasR}. Indeed, from $\delta\left(D_{a}\R_{ABCD}\right)=0$, the $U_{AB}$ component of $\delta(\R_{aAbB}+i\star \R_{aAbB})=0$, and $\delta(\R_{ABCD}+i\star \R_{ABCD})=0$, one finds respectively,
\begin{equation}\label{DRexp}
\begin{aligned}
&     l(l+1) \mathcal{G}_{a} + r^{2}\left(\frac{\mathcal{I}-k}{r^{2}}\right)_{:a} - 2 r \mathcal{S}_{a}{}^{b} r{}_{:b} + r \
    r{}_{:a}{}_ {:b} k{}^{:b} + k{}_{:a} r{}_{:b} r \
    {}^{:b} -  \frac{2 k r{}_{:a} r{}_{:b} r {}^{:b}}{r} - 2 h_{b}{}^{c} r^2 r {}^{:b} r{}_{:a}{}_{:c}  = 0\, , \\ \\
&    -i\mathcal{A}_{a}{}^{c} \epsilon_{bc} + \mathcal{S}_{ab} -  \frac{1}{2} h^{c}{}_{c} r r{}_{:a}{}_{:b} + h_{b}{}^{c} r r{}_{:a}{}_{:c}=0\, , \\ \\
& 4 \mathcal{I} - 4 k + 2 h^{a}{}_{a} r^2 - 4i\mathcal{N} r^2 + 4 k r{}_{:a} r {}^{:a} - 2 h^{b}{}_{b} r^2 r{}_{:a} r {}^{:a}=0\, ,
\end{aligned}
\end{equation}
where we use ``$:$'' to denote covariant differentiation with respect to $\nabla_{a}$. In addition, from $\delta \R^{\alpha}_{\ \mu\alpha\nu}=0$ one has, in particular, that\footnote{Since $\R^{\alpha}_{\ \mu\alpha\nu}=R_{\mu\nu}$ in general, $\delta \R^{\alpha}_{\ \mu\alpha\nu}=0$ is nothing but the usual linearised vacuum Einstein equation whose spherical decomposition can be found in e.g.\cite{Martel:2005ir,Pereniguez:2023wxf}, which indeed imply \eqref{metconst}.} 
\begin{align}\label{metconst}
    h_{\ a}^{a} = 0\, , \quad \nabla_{a} h_{b}^{\ a} = \nabla_{b}\left(\frac{k}{r^2}\right).
\end{align}
Then, recalling the relation \eqref{AlgeBianchiD} and using equations \eqref{DRexp}, one can trade immediately $\mathcal{N},\mathcal{G}_{a},\mathcal{S}_{ab}, \mathcal{A}_{ab}$ by $\mathcal{I},k$ and $h_{ab}$ in \eqref{premast}. Next, after using \eqref{metconst} and the background equations \eqref{bg} one finds that $A_{2}$ in \eqref{premast} vanishes on its own, $A_{2}=0$, while from $A_{1}=0$ one directly gets a remarkably simple, decoupled wave equation,
\begin{equation}\label{RWMasterWaveEqD}
    \left(\Box +\frac{6 M}{r^3}  - \frac{l(l+1)}{r^2} \right)\Psi = 0\, ,
\end{equation}
where $\Box=\nabla^{a}\nabla_{a}$ and we introduced the variable $\Psi$ given by
\begin{equation}
    \Psi \coloneqq \frac{1}{r} \biggl( \mathcal{I} - \frac{4 M k}{r} \biggl)\, .
\end{equation}
Three comments are in order here. First, although in terms of the  curvature component $\mathcal{I}$ our variable $\Psi$ has a simple expression (a fact that simplified drastically the derivation), when $\mathcal{I}$ is written in terms of the metric fluctuation \eqref{met} its form is a rather involved combination,
\begin{equation}\label{Psimet}
    \Psi = \frac{l(l+1)+2}{2r}k- r^{a}\nabla_{a}k+r h_{ab}r^{a}r^{b} - \frac{4 M k}{r^{2}} +i\frac{l(l+1)}{2}r^{3}\epsilon^{ab}\nabla_{a}\left(\frac{j_{b}}{r^{2}}\right)\, ,
\end{equation}
that one would have had to guess if proceeding in the traditional way, using the metric fluctuation instead of the curvature one. Second, since $\Psi$ is a genuine complex quantity (as it contains the component of a complex tensor, $\mathcal{I}$), it encodes two physical degrees of freedom. These are its $+$ and $-$ parts, which can be readily obtained from \eqref{Psimet} and give\footnote{We recall that, since $h_{ab}$, $j_{a}$ and $k$ are components of a real tensor, $h_{\mu\nu}$, one has $h_{ab}^{+}=2h_{ab}$, $j_{a}^{+}=2j_{a}$ and $k^{+}=2k$, while $h_{ab}^{-}=0$, $j_{a}^{-}=0$ and $k^{-}=0$.}
\begin{equation}
    \begin{aligned}\label{Psipm}
    \Psi^{+} &= \frac{l(l+1)+2}{r}k-2 r^{a}\nabla_{a}k+2 r h_{ab}r^{a}r^{b}-\frac{8M}{r^{2}}k\, ,\\
    \Psi^{-} &= i l(l+1)r^{3}\epsilon^{ab}\nabla_{a}\left(\frac{j_{b}}{r^{2}}\right)\, .
\end{aligned}
\end{equation}
Similarly to the electromagnetic case, $\Psi^{\pm}$ consist purely of either even or odd pieces of the metric fluctuation. Since both $\Box$ and the potential in \eqref{RWMasterWaveEqD} are real and independent of the harmonic number $m$, it follows that $\Psi^{-}$ and $\Psi^{+}$ satisfy the \textit{same} equation. In addition to this, they also satisfy the same boundary conditions when considering the associated quasinormal modes, so isospectrality follows as an immediate consequence of \eqref{RWMasterWaveEqD} and \eqref{Psipm}. A third observation is that \eqref{RWMasterWaveEqD} is precisely the Regge--Wheeler equation, which is simpler than Zerilli's, and now it governs both sectors of the fluctuation simultaneously by acting on our complex variable. 

Finally, it is worth relating our variable to those used traditionally in the literature. An immediate identification is that $\Psi^{-}\sim \Psi_{\text{odd}}$, where $\Psi_{\text{odd}}$ is the odd master variable introduced by Cunningham, Price and Moncrief in \cite{CPM} (although we follow the definition in \cite{Martel:2005ir}), which was found to satisfy the Regge-Wheeler equation \eqref{RWMasterWaveEqD}. This is in agreement with our results, although the derivation was very different. It is from the even sector that one can get a non-trivial check of our results. In terms of the Zerilli-Moncrief function $\Psi_{\text{even}}$ introduced in \cite{PhysRevLett.24.737,Moncrief:1974am} (where again we follow the definition of \cite{Martel:2005ir}), we find
\begin{equation}\label{PsiplusZM}
    \Psi^{+}=\frac{l(l+1)}{2}\left[(l-1)(l+2)+\frac{6M}{r}\right] \Psi_{\text{even}} - \frac{6 M }{r^2}k .
\end{equation}
That $\Psi_{\text{even}}$ and $\Psi^{+}$ do not coincide is in agreement with the fact that the Zerilli--Moncrief function satisfies the Zerilli equation, while our variable $\Psi^{+}$ satisfies the Regge--Wheeler one. Thus, a very strong check of our results is to start from the Zerilli equation for $\Psi_{\text{even}}$, rewrite it in terms of $\Psi^{+}$ using \eqref{PsiplusZM}, and use the linearised Einstein equations to show that one arrives at our master equation \eqref{RWMasterWaveEqD}. Although a bit tedious, this computation is straightforward and indeed confirms our result. 

\section{Discussion} \label{Sec3}

We have presented an approach to black hole perturbation theory in spherically symmetric backgrounds based on the so-called curvature wave equations. This addresses simultaneously two conceivably inconvenient aspects of the traditional methods. First, the derivation of decoupled wave equations is direct, and follows from a single component of the curvature wave equation. Second, the formalism encapsulates the degrees of freedom of the even and odd sectors in a single complex variable that satisfies the Regge--Wheeler equation. On the one hand, this implies automatically that the even and odd sectors are isospectral, a fact that is blurred from the original approaches \cite{PhysRev.108.1063,PhysRevLett.24.737} (see also the discussions in \cite{Chandrasekhar:1985kt}). On the other hand, it shows that the dynamics of both sectors is governed by the Regge--Wheeler equation, which is simpler than the original equation derived by Zerilli for the even sector.

More precisely, we considered the radiative modes ($\ell\geq2$) of a metric perturbation,
\begin{equation}
    h=h_{ab}(y)Y(z)dy^{a}dy^{b}+2j_{a}(y)X_{A}(z)dy^{a}dz^{A}+k(y) U_{AB}(z)dz^{A}dz^{B}\, ,
\end{equation}
which, in turn, induce fluctuations of the self-dual curvature tensor \eqref{CompR}, whose purely spherical component has the form
\begin{equation}
    \delta \R_{ABCD}=\mathcal{I}(y)Y(z)\epsilon_{AB}\epsilon_{CD}\, .
\end{equation}
Then we showed that, for vacuum fluctuations about Schwarzschild's solution, the spherical component of the curvature wave equation $\epsilon^{AB}\epsilon^{CD}\left(\delta \triangle\R\right)_{ABCD}=0$ reduces to the Regge--Wheeler equation \eqref{RWMasterWaveEqD} for the variable
\begin{equation}
    \Psi= \frac{1}{r} \biggl( \mathcal{I} - \frac{4 M k}{r} \biggl)\, .
\end{equation}
Furthermore, its real and imaginary parts (or more precisely its $+$ and $-$ parts, see \eqref{pm}) contain respectively only even and odd pieces of the metric fluctuation, as shown in \eqref{Psipm}, which together with equation \eqref{RWMasterWaveEqD} automatically imply isospectrality between sectors, as advertised.

While our approach and results are new, it is worth noticing that some works in the literature obtained conclusions along lines similar to ours. In particular, employing the Newman--Penrose formalism reference \cite{Aksteiner:2010rh} found a complex master variable whose real and imaginary parts consisted solely of even and odd pieces of the metric fluctuation, and that satisfied the Zerilli and Regge--Wheeler equation, respectively. Also from a different approach, references \cite{Sasaki1981TheRE,Chaverra:2012bh} found a variable in the even sector that satisfies the Regge--Wheeler equation, instead of Zerilli's. In some sense, our formalism seems to synthesize both facts using a decomposition into spherical harmonics, and it would be interesting to study closer the connections among these approaches. It would also be interesting to classify these variables according to the scheme in \cite{Lenzi:2021wpc}. The tasks of reconstructing the metric from our variable, as well as studying in detail the asymptotic radiation fields will be considered elsewhere.

There are some natural continuations of our work. First, the ideas described in Sections \ref{Sec1} and \ref{Sec2} can be extended easily to include matter. Given that we obtained simpler equations for the even sector from this approach, one may have a similar simplification beyond vacuum. This could allow the treatment of the even sector in some spacetimes which, due to their intricate structure, have only been considered in the odd sector (see e.g. \cite{Bamber:2021knr,Redondo-Yuste:2023ipg}). It would also be interesting to account for a cosmological constant, and study the consequences on isospectrality in that case \cite{Cardoso:2001bb}. A second natural extension is to consider higher orders in perturbation theory. As discussed in the introduction, the structure of linear perturbation theory is nested in all next orders. In our case, the formalism allows one to use a single propagator (instead of one for each sector), and gives a prescription to define the master variable at any given order, as $\Psi^{(n)}=r^{-1}(\mathcal{I}^{(n)}-4M k^{(n)}/r)$. The component $\mathcal{I}^{(n)}$, which is geometrically defined as $\delta^{(n)}\R_{ABCD}=\mathcal{I}^{(n)}Y\epsilon_{AB}\epsilon_{CD}$, includes not only $n$-th order metric perturbations, but also a specific combination of products of lower order ones. It would be interesting to compute the higher-order source terms of this particular variable, and whether this simplifies their process of renormalisation \cite{Brizuela:2009qd}. In addition, to address higher-orders one often needs to reconstruct the entire metric in a particular gauge, so it would be interesting to explore the structure of reconstruction operators in our formalism. Finally, a variant of the approach presented here might be useful in higher-dimensional backgrounds as long as these keep the 2-sphere factor in the metric (e.g. $6D$ single-spin Myers--Perry black holes \cite{Myers:1986un}). In that case, one should work with $R_{\mu\nu\rho\sigma}$, perform a similar spherical expansion as in \eqref{RiemannHarmonicProjD} and subsequently compute $\epsilon^{AB}\epsilon^{CD}\delta\left(\triangle R\right)_{ABCD}=0$. Finally, it would also be interesting to study the slow-spin expansion proposed in \cite{Franchini:2023xhd} in terms of our formalism, and check whether isospectrality remains manifest at higher-orders in the spin.

\section*{Acknowledgements}

We thank José Beltrán, Emanuele Berti, Vitor Cardoso, Francisco Duque, Nicola Franchini, Michele Lenzi, Marc Mars, Adam Pound, Jaime Redondo--Yuste, Luca Santoni, Carlos Sopuerta and Nick Speeney for useful conversations. We also thank Vitor Cardoso, Jaime Redondo--Yuste, Luca Santoni, Carlos Sopuerta and Nick Speeney for feedback on the manuscript. We are deeply grateful to Eric Poisson for his detailed correspondence and valuable discussions on earlier drafts of this work. We used substantially the \texttt{xPert} and \texttt{xTensor} packages of \texttt{xAct}, and our notebooks are available upon request. We acknowledge financial support by the VILLUM Foundation (grant no. VIL37766)
and the DNRF Chair program (grant no. DNRF162) by
the Danish National Research Foundation. This project
has received funding from the European Union’s Horizon 2020 research and innovation programme under the
Marie Sklodowska-Curie grant agreement No 101007855
and No 101007855. 

\appendix
\section{Explicit Expression for $\epsilon^{AB}\epsilon^{CD}\delta\left(\triangle \R\right)_{ABCD}=0$}\label{App}

To linearise \eqref{GravCWE} in terms of $\delta \R$, one has to convert the normal Riemann tensor terms into their complex counterparts. To convert the second term in \eqref{GravCWE} one can use the self-dual condition \eqref{DR} to show $R_{\mu \nu}^{\ \ \gamma \lambda}\mathbb{R}_{\gamma \lambda \rho \sigma} = \frac{\mathbb{R}_{\mu \nu}^{\ \ \gamma \lambda}\mathbb{R}_{\gamma \lambda \rho \sigma}}{2}$. The remaining Riemann terms can be converted using $R_{\mu \nu \rho \sigma} = \frac{\mathbb{R}_{\mu \nu \rho \sigma} + \bar{\mathbb{R}}_{\mu \nu \rho \sigma}}{2}$, denoting complex conjugation with a bar $-$ on top. We find more convenient to work with $\star \triangle \R=0$ instead of $\triangle\R=0$, which reads
\begin{equation}
    \label{ModGravCWE}
    \Box \mathbb{R}_{\mu \nu \rho \sigma} + \frac{1}{2} \mathbb{R}_{\rho \sigma \lambda \delta} \mathbb{R}_{\mu \nu}^{\ \ \lambda \delta} - \frac{i}{2} \epsilon_{\rho \sigma}^{\ \ \gamma_1 \gamma_2} \Biggl( 2 \mathbb{R}^{\lambda}_{\ \ \mu \gamma_1 \delta} \mathbb{R}_{\lambda \nu \gamma_2}^{\ \ \ \ \delta} + \Bar{\mathbb{R}}^{\lambda}_{\ \ \mu \delta \gamma_2} \mathbb{R}_{\lambda \nu \gamma_1}^{\ \ \ \ \delta}  + \Bar{\mathbb{R}}^{\lambda}_{\ \ \nu \delta \gamma_2} \mathbb{R}_{\mu \lambda  \gamma_1}^{\ \ \ \ \delta}\Biggl)=0\, ,
\end{equation}
and consider its linearisation $\delta\left( \star \triangle \R\right)=0$. Of course, since $\R+i\star \R=0$ implies $\triangle \R+i\star \triangle \R=0$, one has that $\delta \left(\triangle \R\right)=-i\delta\left(\star \triangle \R\right)$, and proceeding this way we find 
\begin{equation}
\begin{aligned}\label{premastexp}
&\epsilon^{AB} \epsilon^{CD} \left(\delta \triangle \R\right)_{ABCD}= \\
 &\Biggl\{  
 -  \frac{24 r{}_{:a} \mathcal{I} {}^{:a}}{r}+ 4 \mathcal{I} {}^{:a}{}_{:a}+\left[-4\frac{l(l+1)-2}{r^{2}}-\frac{16 r{}^{:a}{}_{:a}}{r}  +  \frac{24  r{}_{:a} r{}^{:a}}{r^2}\right]\mathcal{I}+8i\left(1 -  r{}_{:a} r{}^{:a}\right) \mathcal{N} 
\\
&-8\left(1-r{}_{:b} r{}^{:b}\right)\tilde{k}{}^{:a}{}_{:a}+8\left[3\frac{r_{:a}}{r}\left(1-r{}_{:b} r{}^{:b}\right)+r{}_{:a}{}_{:b} r{}^{:b}\right]\tilde{k}{}^{:a}
\\
&+8\left[\frac{l(l+1)\left(1-r{}_{:a} r{}^{:a}\right)+(3r{}_{:b} r{}^{:b}-2)r{}_{:a} r{}^{:a}-1}{r^{2}}+2\frac{r{}^{:a}{}_{:a}\left(1-r{}_{:b} r{}^{:b}\right)-2 r{}^{:a} r{}_{:a}{}_{:b} r{}^{:b}}{r}\right]\tilde{k}
\\
&-16 \frac{l(l+1)}{r}\mathcal{G}^{a} r{}_{:a}+ \left(16  r{}^{:a} r{}^{:b} - 4 r  r{}^{:a}{}^{:b}\right)\mathcal{S}_{ab} - 12i \mathcal{A}^{ab} \epsilon_{a}{}^{c} r r{}_{:b}{}_{:c}-16 i l(l+1)\tilde{j}^{a} r{}^{:b} r{}_{:c}{}_{:(a}\epsilon_{b)}{}^{c}
 \\
&-4 r\left[ r{}^{:a}\left(1-r{}_{:c} r{}^{:c}\right)+r r{}^{:c}  r{}_{:c}{}^{:a}\right]\tilde{h}^{b}{}_{b}{}_{:a}+8 r\left[ r{}^{:a}\left(1-r{}_{:c} r{}^{:c}\right) + r r{}^{:c} r{}_{:c}{}^{:a}  \right] \tilde{h}_{a}{}^{b}{}_{:b}
\\
&+4\left[-1+r{}_{:b} r{}^{:b}\left(2-r{}_{:c} r{}^{:c}\right)-r^2 r{}_{:b}{}_{:c} r{}^{:b}{}^{:c}\right]\tilde{h}^{a}{}_{a}
\\
&+8\left[\left(r{}^{:a} r{}^{:b}+rr{}^{:a}{}^{:b}\right)\left(1-r{}_{:c} r{}^{:c}\right)+r r{}^{:c} \left(r r{}_{:c}{}^{:b}{}^{:a}+2 r{}^{:a} r{}_{:c}{}^{:b}\right)\right]\tilde{h}_{ab}\Biggr\} Y
\\
&-\Biggl\{4 r \bar{\mathcal{S}}^{ab} r{}_{:a}{}_{:b} + 4i
\bar{\mathcal{A}}^{ab} \epsilon_{a}{}^{c} r r{}_{:b}{}_{:c} \Biggr\} \bar{Y}, 
\end{aligned}
\end{equation}
that give the functions $A_{1}$ and $A_{2}$ in \eqref{premast} (recall that ``$:$'' stands for covariant differentiation with respect to $\nabla_{a}$). Writing \eqref{premastexp} in terms of $\mathcal{I}$, $k$ and $h_{ab}$ only, by using \eqref{DRexp} and \eqref{metconst}, one indeed finds that $A_{2}$ vanishes independently, $A_{2}=0$, and $A_{1}=0$ gives \eqref{RWMasterWaveEqD}.

\bibliography{[JCAP]Bibliography.bib}

\end{document}